\documentclass[10pt,conference]{IEEEtran}
\IEEEoverridecommandlockouts
\usepackage{comment}
\usepackage{cite}
\usepackage{amsmath,amssymb,amsfonts}
\usepackage{algorithmic}
\usepackage{graphicx}
\usepackage{textcomp}
\usepackage{xcolor}
\def\BibTeX{{\rm B\kern-.05em{\sc i\kern-.025em b}\kern-.08em
    T\kern-.1667em\lower.7ex\hbox{E}\kern-.125emX}}

\usepackage{fancyhdr}

\newcommand{\fapesp}{\includegraphics[scale=0.035]{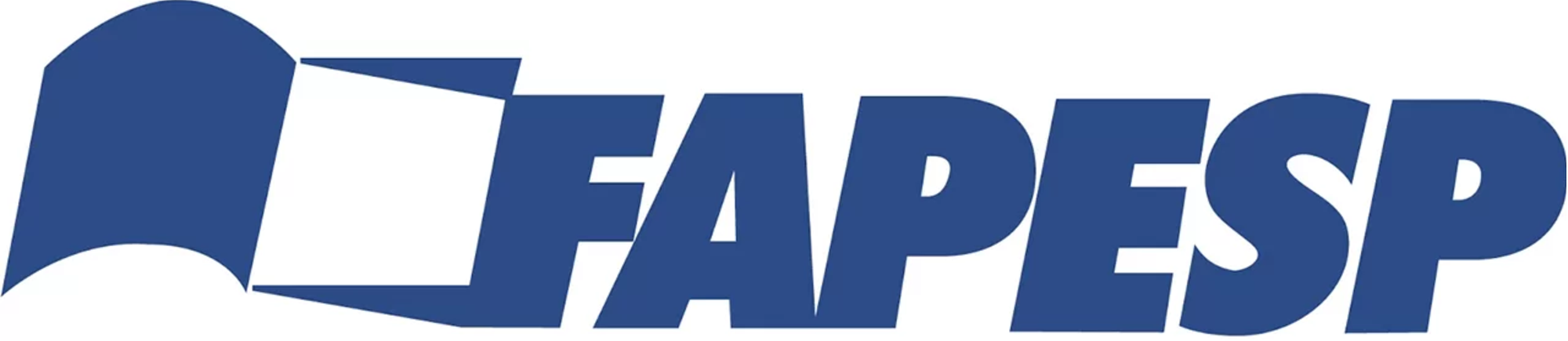}}
\newcommand{\smartness}{\includegraphics[scale=0.158]{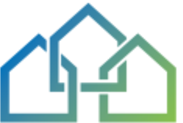}}

\begin{document}

\title{POSMAC: Powering Up In-Network AR/CG Traffic Classification with Online Learning}

\begin{comment}
\footnotesize {\textsuperscript{*}Note: Sub-titles are not captured for https://ieeexplore.ieee.org  and should not be used}
\thanks{Identify applicable funding agency here. If none, delete this.}
\end{comment}

\author{\IEEEauthorblockN{Alireza Shirmarz, Fábio Luciano Verdi}
\IEEEauthorblockA{\textit{Dept. of Computer Science} \\
\textit{Universidade Federal de São Carlos (UFSCar)}\\
Sorocaba, Brazil \\
{ashirmarz@ufscar.br, verdi@ufscar.br}}
\and
\IEEEauthorblockN{Suneet Kumar Singh, Christian Esteve Rothenberg}
\IEEEauthorblockA{\textit{Dept. of Computer Engineering} \\
\textit{Universidade Estadual de Campinas (UNICAMP)} \\
Campinas, Brazil \\
{ssingh@dca.fee.unicamp.br, chesteve@dca.fee.unicamp.br}}

%\and
%\IEEEauthorblockN{Given Name Surname}
%\and
%\and
}

\maketitle

%\marginnote{\textit{Accepted at **IEEE Conference on NFV-SDN 2024**, to be indexed by IEEE Xplore.}}
\fancypagestyle{plain}{
    \fancyhf{} % Clear default header/footer
    \fancyfoot[C]{Accepted at IEEE Conference on Network Function Virtualization and Software Defined Networks (NFV-SDN) 2024.}
    \renewcommand{\headrulewidth}{0pt}  % Remove header line
    \renewcommand{\footrulewidth}{0pt}  % Remove footer line
}

\pagestyle{plain}  % Apply the style to all pages

\begin{abstract}
In this demonstration, we showcase POSMAC\footnote{Platform of Optimization \& Deployment of the Online Self-Trainer Model for AR/CG Traffic Classification.}, a platform designed to deploy Decision Tree (DT) and Random Forest (RF) models on the NVIDIA DOCA DPU, equipped with an ARM processor, for real-time network traffic classification. Developed specifically for Augmented Reality (AR) and Cloud Gaming (CG) traffic classification, POSMAC streamlines model evaluation, and generalization while optimizing throughput to closely match line rates. 
\end{abstract}

\begin{IEEEkeywords}
ML model deployment, Online traffic classification, AR/CG, retraining platform.
\end{IEEEkeywords}

\section{Introduction}
Network traffic classification is crucial in applications like AR and CG, where encryption and dynamic port usage obscure traditional identifiers~\cite{b31}. In our previous work, we utilized RTP features to achieve a 94.8\% accuracy in classifying traffic~\cite{netsoft2024}. To further enhance model accuracy, we developed POSMAC\footnote{https://github.com/dcomp-leris/POSMAC.}, a platform that refines models by integrating new PCAP data and employing transfer learning techniques. This pre-deployment process, essential for avoiding the costs of deploying unoptimized models, outputs a robust and well-generalized model. 
%Leveraging NVIDIA DOCA on BlueField-3 DPU within a Docker environment, POSMAC ensures models are efficient and cost-effective to deploy, with the final model ready for deployment on the dataplane as suggested by \cite{Myref2:leo2024,Myref1:IIsy2024,Myref4:jewel2024, myref6:swamy2023homunculus}. This approach maximizes throughput and performance in real-time traffic classification scenarios.
POSMAC, leveraging NVIDIA DOCA on BlueField-3 DPU in a Docker environment, produces models that are efficient and cost-effective for deployment on BlueField-3 and other dataplanes, as supported by \cite{Myref2:leo2024,Myref1:IIsy2024,Myref4:jewel2024, myref6:swamy2023homunculus}. 

%Network traffic classification is crucial when encryption or dynamic port usage obscures IP and port addresses~\cite{b31}, particularly in AR and CG applications which are rapidly expanding~\cite{netsoft2024}. Operators must classify traffic based on packet features since traditional identifiers are often unavailable due to their dynamic nature and encryption. Despite some inherent uncertainties in this process, the accuracy of classifications improves with the use of larger datasets. Furthermore, training of machine learning models is an ongoing process; therefore, an adaptable platform is essential to facilitate continuous retraining and updating of models to keep pace with evolving network patterns.
%The challenge extends to deploying these models on programmable hardware to classify traffic at line rates~\cite{Myref2:leo2024,Myref1:IIsy2024,Myref4:jewel2024, myref6:swamy2023homunculus}. Our platform, POSMAC, leverages NVIDIA DOCA emulator on the BlueField 3.0 with an ARM processor within a Docker Container Engine environment. This setup allows for efficient model deployment and high packet processing rates through the use of multiprocessing cores. Unlike previous approaches that focus on serializing and offloading models to the data plane, POSMAC enhances model training, generalization, and resource efficiency prior to hardware deployment. This method not only simplifies model evaluation and generalization but also aims to maximize throughput to reach line rates using the DPU on the DOCA.

\section{Architecture}
POSMAC is a Docker-based platform consisting of four main components: (1) the Pcap Pool, (2) Traffic Classifier (TC), (3) Application Servers (APS), and (4) Online Trainer (OT), as illustrated in Fig 1. Each component is labeled and marked with green circles, connected by directed grey arrows indicating communication direction. Communication occurs at the datalink layer using MAC addresses, with dedicated networks ensuring exclusive connectivity between components (shown by the grey arrows). The online learning data flows, depicted by blue, red, and black dotted lines, indicate data transfer direction across the platform. Detailed descriptions of each component follow in subsequent sections.
% Architecture
\begin{figure*}[tp]
\centering
\includegraphics[trim=0.0cm 0.0cm 0.0cm 0.0cm, width=0.75\textwidth]{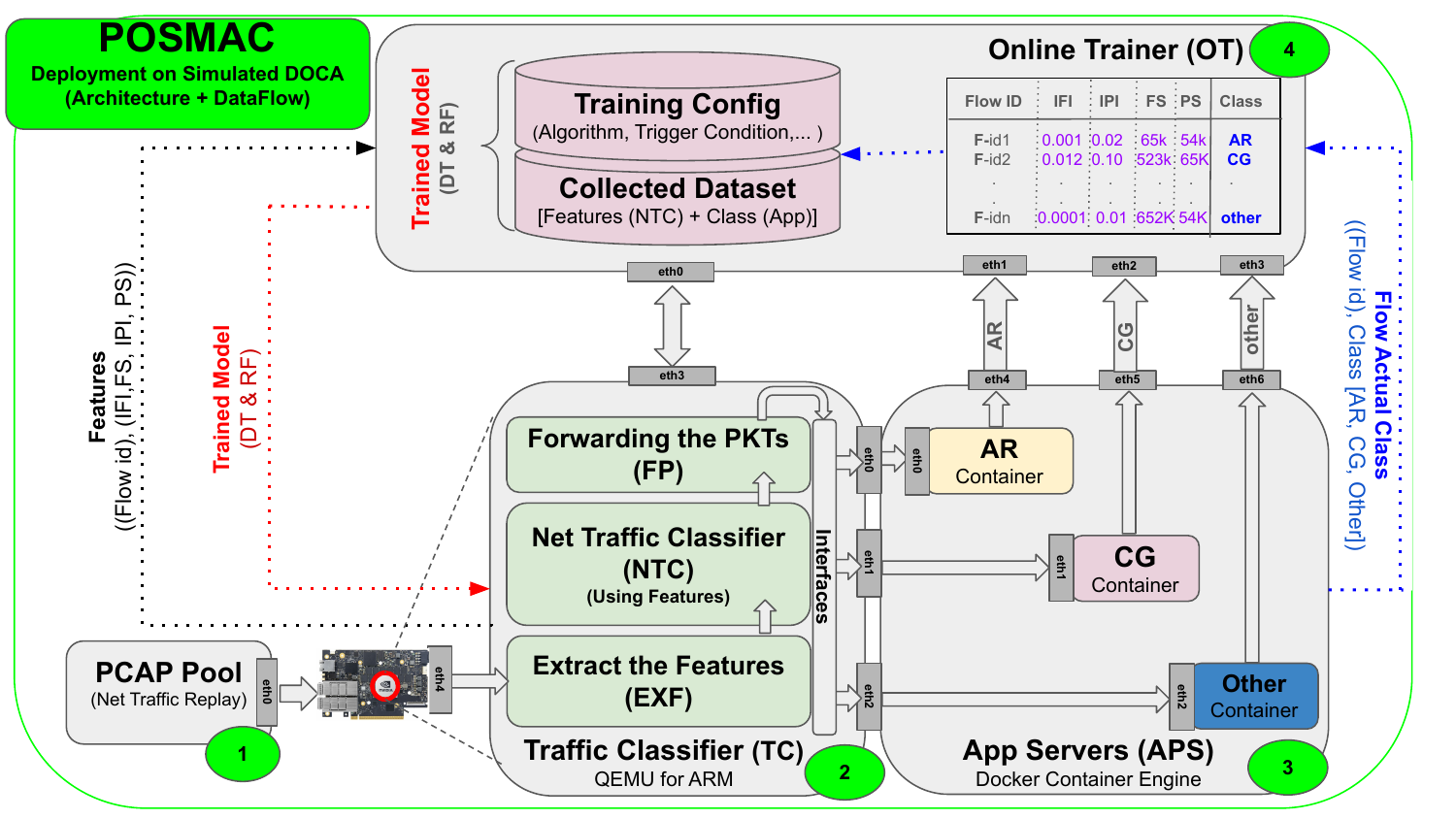} % Adjust the height as needed
\caption{POSMAC Architecture: Dotted lines illustrate online learning data flows — black for feature extraction, blue for class labeling, and red for transferring the retrained ML model. \label{fig:arch}}
 \vspace{-0,40cm}
\end{figure*}

\subsection{PCAP Pool}\label{AA}
The PCAP Pool is a key component of the POSMAC platform, simulating incoming network traffic to the server. It stores pre-collected PCAP files from various applications, with traffic categorized into three labels: `AR', `CG', and `other', representing AR, CG, and other UDP-based applications like video conferencing and live streaming. After preprocessing with TCPReplay to adjust data rates, traffic is routed from the PCAP Pool to the TC component, where it is efficiently classified. The PCAP Pool can simulate traffic or be replaced by real user-generated traffic, enabling testing under diverse frame rates, resolutions, and network bandwidths.

\subsection{Traffic Classifier (TC)}
The Traffic Classifier component within the POSMAC platform utilizes pre-trained models, such as DT or RF, initially deployed on the Data Processing Unit (DPU) BlueField 3.0, which features an ARM processor.
These models are iteratively trained and improved using the OT block. This component receives network packets and routes them to the appropriate server based on the classified application type. It directly connects to three distinct application servers—each representing a specific class (AR, CG, or other applications)—to ensure targeted traffic handling. The TC comprises three key operations: \textbf{(1) Extracting Features (EXT):} This function extracts relevant features from the packets. \textbf{(2) Network Traffic Classifier (NTC):} Utilizes the RF or DT model to classify network flows using the extracted features. \textbf{(3) Forwarding Packets (FP):} Routes packets to the corresponding application server based on their classified type.
Additionally, this component maintains a direct link to the OT, periodically sending batched feature data to facilitate continuous model training and refinement.

\subsection{Application Servers (APS)}
In the POSMAC platform, the Application Servers component deploys servers with heterogeneous computing resources, which may be physical, virtual, or containerized. Each server type connects directly to both the TC and OT components. The platform designates one container per application class— specifically AR, CG, and others. Each application server, such as the AR server, receives network flows and verifies their correct routing based on the server IP configurations set in the PCAP Pool. It also labels the flows accordingly. $Flow_{ID}$s, which are unique identifiers, along with their corresponding labels (AR, CG, or other), are then transmitted to the OT. This connection to the OT is facilitated directly through a designated NIC, ensuring efficient communication and data handling within the platform.

\subsection{Online Trainer (OT)}
The OT is a crucial component that periodically trains the model, maintaining direct connections to both the TC and the Apps. It collects datasets from the TC's EXT function, which includes Packet Size (PS), Inter Packet Interval (IPI), Frame Size (FS), and Inter Frame Interval (IFI), each tagged with a unique $flow_{ID}$. These datasets are further enriched with labels provided by the Application Servers, corresponding to each $flow_{ID}$.
The OT is programmed with configurable rules that dictate the timing and conditions under which training occurs. Once the model is updated and enhanced with new data, it is serialized and transmitted back to the TC, where it is loaded and deployed according to predefined configuration rules. This cyclical process ensures that the classification model remains accurate and up-to-date with the latest network traffic patterns.

\section{Current Deployment \& Output}
Our initial deployment utilizes DT and RF models to classify network traffic into categories such as AR, CG, and others, based on metrics like PS, IPI, FS, and IFI. 
Using the DOCA simulator for BlueField 3.0 DPU, our models initially achieve 95.7\% accuracy, which approaches nearly 100\% through transfer learning as the model is retrained with new data. The improvement rate depends on the retraining strategy in OT. 
Our platform not only provides real-time reports on accuracy and resource usage but also enables detailed analysis of TC performance, including packet processing rates and system requirements. 
The results are carefully logged to optimize classification accuracy and throughput. In the demo, we will showcase POSMAC components by analyzing the accuracy, throughput, and latency of the TC as it processes traffic from various applications using pre-collected PCAP files.
Our next phase is to deploy these models on BlueField 3.0 hardware to validate and enhance performance in real-world conditions. 
The primary challenge in deploying online learning models on hardware is obtaining accurate, continuously updated class labels in real-world environments. However, the platform uses online learning to enhance and generalize models before hardware deployment, though this real-world loop poses the challenge of evolving labels.
\vspace{-0,10cm}
\section*{Acknowledgment}
This work was supported by Ericsson Telecomunicações Ltda., and by the Sao Paulo Research Foundation (FAPESP), \fapesp~grant \texttt{2021/00199-8}, CPE SMARTNESS~\smartness.
%This study was partially funded by \textit{CAPES}, Brazil - Finance Code \texttt{001}.
%This work was supported by Ericsson Telecomunicações Ltda., and by the Sao Paulo Research Foundation (FAPESP), \fapesp~grant \texttt{2021/00199-8}, CPE SMARTNESS~\smartness. 
\bibliographystyle{IEEEtran}
\bibliography{reference} % Assumes you have a myreferences. bib file
\begin{comment}
\vspace{12pt}
\color{red}
IEEE conference templates contain guidance text for composing and formatting conference papers. Please ensure that all template text is removed from your conference paper prior to submission to the conference. Failure to remove the template text from your paper may result in your paper not being published.

\end{comment}
\begin{comment}
\section{Demo Technical Requirements}
\begin{itemize}
    \item Single laptop with Ubuntu Linux OS
    \item Power \& Internet Access 
    \item Setup Time is around 30 min
    \item The Space provided by NFV\_SDN
\end{itemize}
\end{comment}

\end{document}